# Additional experimental evidence against a solar influence on nuclear decay rates

Eric B. Norman

Dept. of Nuclear Engineering

Univ. of California

Berkeley, CA 94720

A new paper by Jenkins *et al*. claims additional evidence for a solar influence on nuclear decay rates (http://arxiv.org/abs/1207.5783). The data shown in this paper comes from a Geiger-Muller detector at Ohio State University (OSU) designed for radiation safety measurements of radioactive swipes. The data shown in Fig. 1 of this new paper appears to show temporal variations. However, the authors claim that the $^{36}$Cl counting rate in this OSU data set is higher in the winter and lower in the summer. This is **opposite** to the effect seen in the BNL experiment (Alburger *et al.,* Earth & Planet. Sci. Lett. **78** (1986) 168, that also used $^{36}$Cl) to which Jenkins *et al.* refer. If these variations were really due to some phenomenon related to the distance between the Earth and the Sun, it can't produce an effect of one sign in one experiment and the opposite sign in another. Furthermore, the amplitude of the variations observed in the OSU data is at least a factor of five larger than that seen in the BNL experiment.

What Jenkins *et al*. refer to as the evidence for annual variations with a maximum in winter in the BNL data is the ratio of $^{32}$Si/$^{36}$Cl shown in Figures 2 and 4 of Alburger *et al.* The actual $^{36}$Cl counting rate measured in the BNL experiment is shown in Figure 3 of Alburger *et al*. In the BNL data, the $^{36}$Cl reference counting rate was observed to be **lower** in the winter and **higher** in the summer. This can be seen most clearly in the data taken between years 2 and 4 shown in Figure 3. The authors of the BNL paper noticed that this annual variation was in phase with the annual variation in the average temperature in the New York area. They noted "since lower temperature and lower humidity both result in higher air density at constant pressure, the lower-energy $^{36}$Cl β rays would be absorbed more effectively than those of $^{32}$P and one would expect the $^{32}$Si/$^{36}$Cl ratio to be **higher** in the winter, as observed". Although they could not quantitatively explain the magnitude of the effect they observed, these authors did point out this interesting coincidence.

The 180-degree phase difference between the $^{36}$Cl counting rates measured at OSU and BNL and their large difference in amplitudes point to systematic effects associated with the measuring apparatus rather than any dependence on the distance between the Earth and the Sun. Nevertheless, because of the potential implications for geochronology, archeology, and other sciences, carefully controlled experiments dedicated to searching for temporal variations in nuclear decay rates are still warranted.